\documentclass[manuscript]{aastex}

\slugcomment{Not to appear in Nonlearned J., 45.}

\shorttitle{HE0047 and SDSS1155}
\shortauthors{Rojas et al.}

\begin{document}

\title{Strong Chromatic Microlensing in HE0047-1756 and SDSS1155+6346}

\author{K. Rojas, V. Motta}
\affil{Instituto de F\'isica y Astronom\'ia, Universidad de Valpara\'iso, Avda. Gran Breta\~na 1111, Playa Ancha, Valpara\'iso 2360102, Chile}
\email{karina.rojas@uv.cl, veronica.motta@uv.cl}

\author{E. Mediavilla}
\affil{Instituto de astrof\'isica de Canarias, Avda. V\'ia Lactea s/n, La Laguna, Tenerife 38200, Spain}
\affil{Departamento de astrofisica, Universidad de La Laguna, La Laguna, Tenerife 38205, Spain}
\email{emg@iac.es}


\author{E. Falco}
\affil{Whipple Observatory, Smithsonian Institution, 670 Mt. Hopkins Road, P.O. Box 6369, Amado, Arizona 85645, USA}
\email{falco@cfa.harvard.edu}

\author{J. Jim\'enez-Vicente}
\affil{Departamento de F\'isica Te\'orica y del Cosmos, Universidad de Granada, Campus de Fuentenueva, 18071
Granada, Spain}
\affil{Instituto Carlos I de F\'isica Te\'orica y Computacional, Universidad de Granada, 18071 Granada, Spain}
\email{jjimenez@ugr.es}

\author{J. A. Mu\~noz}
\affil{Departamento de Astronom\'ia y Astrof\'isica, Universidad de Valencia, 46100 Burjassot, Valencia, Spain}
\email{jmunoz@uv.es}

\begin{abstract}

We use spectra of the double lensed quasars HE0047-1756 and SDSS1155+6346 to study their unresolved structure through the impact of microlensing. There is no significant evidence of microlensing in the emission line profiles except for the Ly$\alpha$ line of SDSS1155+6346, which shows strong differences in the shapes for images A and B. However, the continuum of the B image spectrum in SDSS1155+6346 is strongly contaminated by the lens galaxy and these differences should be considered with caution. Using the flux ratios of the emission lines for image pairs as a baseline to remove macro-magnification and extinction, we have detected strong chromatic microlensing in the continuum measured by CASTLES\footnote{www.cfa.harvard.edu/castles/}, in both lens systems, with amplitudes $0.09 (\lambda 16000) \lesssim |\Delta m |\lesssim  0.8 (\lambda 5439)$ for HE0047-1756, and $0.2 (\lambda 16000) \lesssim |\Delta m |\lesssim  0.8 (\lambda 5439)$ for SDSS1155+6346. Using magnification maps to simulate microlensing, and modeling the accretion disk as a Gaussian source (I $\propto$ exp(-R$^{2}$/2r$^2_s$)) of size r$_{s}$ $\propto$ $ \lambda^{p} $ we find, r$_{s}$ = 2.5$_{-1.4}^{+3.0}$ $\sqrt{M/0.3M_{\odot}}$ light days and p = 2.3 $\pm$ 0.8, at the rest frame for $\lambda$ = 2045, for HE0047-1756 (log prior), and r$_{s}$ = 5.5$_{-3.3}^{+8.2}$ $\sqrt{M/0.3M_{\odot}}$ light days and p = 1.5 $\pm$ 0.6 at the rest frame of $\lambda$ = 1398, for SDSS1155+6346 (log prior). Contrary to other studied lens systems, the chromaticity detected in HE0047-1756 and SDSS1155+6346 is large enough to fulfill the thin disk prediction. The inferred sizes, however, are very large compared to the predictions of this model, especially in the case of SDSS1155+6346. 


\end{abstract}

\keywords{gravitational lensing: strong, microlensing, quasar, accretion disks. Individual: HE0047-1756, SDSS1155+6346}



\section{Introduction}

Gravitational lens systems are a powerful tool to study lens galaxies \citep{ko01,ogu02,dav03,mou03,man09}, to resolve lensed quasar structure \citep{poo07,Mos2009,black2011,Mos2011,Mu2011,Med2011,gue2013a}, and to estimate cosmological constraints \citep{sch92,koop99,ogu07,ju2010,bal13}. Since the discovery of lensed quasars, anomalies in the flux ratios between images compared to the predictions of otherwise reliable models were found. These anomalies were thought to be associated with different phenomena: a complex mass distribution in the lens galaxy, dust extinction, dark matter substructure, and microlensing \citep{ko91,con05,yon08}. One of the most likely explanations,  quasar microlensing, is produced by compact objects in the lens galaxy \citep{chan&ref79,wamb06}. Two observational methods either based on the use of light curves or spectra have been used to study this effect. The first one uses the variability rate observed in the light curves to determine the time delay and measure the microlensing magnification \citep{ra91,yon98,poin10}. The disadvantage of this method is the need for monitoring over periods of years (for most systems microlensing variability is expected over scales greater than 10 years \citep{ko04}). The second method uses the magnitude difference between continuum and emission lines in a single-epoch spectrum to measure microlensing \citep{Med2011,Mot12}. The disadvantage of this method is the lack of a time-delay correction ($\sim$ 30 days for an image separation of 1 arcsec \citep{yon08}) to untangle microlensing from intrinsic variability.

Microlensing is size sensitive. Regions of the source with a size comparable to the Einstein radius of the microlenses or smaller are significantly magnified \citep{wamb98,wamb06,schm&wamb2010}. Then, because the sizes of the emitting regions vary with wavelength, chromatic microlensing is expected \citep{wamb&pacs91,wit95,Mos2009,Med2011}. The effect is stronger for shorter wavelengths \citep[e.g X-rays,][]{poo07}, and can be neglected in the IR \citep[e.g][]{poin07}. 

Intrinsic quasar variability together with time delay could mimic chromatic microlensing \citep{yon08,Mot12}. We estimate this effect for both systems, following \cite{yon08} recipe. We assume an intrinsic magnitude of M$_{I}$ = -21 for the quasar and we estimated the time delay modeling a SIS using \textit{lensmodel} \citep{keet01}. For HE0047-1756, with a time delay of $\sim$ 32.9 days, the expected intrinsic variability is $\lesssim$ 0.08 mag, and the chromaticity change is $\lesssim$ 0.03 mag. For SDSS1155+6346, with a time delay of $\sim$ 11.2 days, the expected intrinsic variability is $\lesssim$ 0.06 mag, and the chromaticity change is $\lesssim$ 0.02 mag. The chromaticity observed for both systems is at least one order of magnitude larger than expected from intrinsic variability.

The BLR has a size of $\sim$60 light-days \citep{bentz09,zu2011,gue2013a} which is much larger than the size of $\sim$4 light-days of the accretion disk (\cite{jv2012,jv2014} and reference therein). It is therefore clear that the broad emission lines are expected to be much less affected by microlensing than the continuum. However there are examples of such variation due to microlensing. For instance, microlensing could affect the broad wings of the high ionization broad emission line profiles \citep{gue2013a,gue2013b} but it is expected to affect very weakly the cores of the lines and the low ionization lines \citep{pop01,abj02,ric04,lew&iba04,gom06,gue2013a}. Thus, following the steps of \cite{Mot12} (see also \cite{Med2011} and references therein) we use the images (A and B) of double lensed quasars to estimate the amplitude of microlensing in the continuum from the magnitude difference of the continuum adjacent to the emission lines, $(m_{B}-m_{A})_{cont.}$, taking as baseline the magnitude differences of the emission line cores, $(m_{B}-m_{A})_{core}$, to remove the effects of macro-magnification and extinction, $\Delta m=(m_B-m_A)_{cont.}-(m_B-m_A)_{core}$. We apply this analysis to spectra of HE0047-1756 and SDSS1155+6346 with the aim of using the variation in the microlensing amplitude with wavelength to constrain the size and the temperature profile of each emitting region.

The structure of the paper is as follows. In section 2 we present the data. In section 3 we give details about the procedure to determine the continuum and line core emission and about the Bayesian analysis used to estimate the size of the accretion disk and the slope of the temperature profile. We discuss the results in section 4, and present the conclusions in section 5.

\section{Observations and Data Reduction}

We obtained spectra for two lens systems: we observed HE0047-1756 with the IMACS Long-Camera at the Magellan telescope in 2008 with a seeing of 0.61$"$, and SDSS1155+6346 with the Blue Channel spectrograph at the MMT in 2010 with a seeing of 0.7$"$. The  wavelength range, spectral resolution and dispersion of the spectra used are 3650-9740 \AA, 6.75 \AA, and 0.743 (\AA/pix), 3000-10000 \AA, 6.47 \AA, 1.96 (\AA/pix), for Magellan and MMT respectively. The position angle of the slit was chosen to observe the two lensed quasar images at once. Table \ref{tab1} shows the log of our observations.

The data reduction was performed with IRAF, and includes bias subtraction, flat normalization and wavelength calibration. As the separation between the components is small (1.43$"$ for HE0047-1756 and 1.94$"$ for SDSS1155+6346\footnote{astrometic data obtained from CASTLES}) the spectra slightly overlap, and the flux extraction of each component was made fitting a Gaussian to each component through each column of the 2D spectra (columns correspond to wavelength). The separation between Gaussians was fixed using the images positions in CASTLES. We did not flux calibrate our data  because we are interested only in the flux ratio between components. We used data from CASTLES in three different bands (F160W, F555W, F814W) acquired with HST in 2003, and additional data from the literature \citep{slu2012,pin04}.

\section{Method}

As discussed in Section 1, the method we use to measure microlensing is based on the comparison between the continuum and the emission line flux ratios, $\Delta m=(m_B-m_A)_{cont.}-(m_B-m_A)_{core}$. We used DIPSO in STARLINK to fit the continuum on either side of the emission lines with a line (flux = a$\lambda$ + b). After subtracting the continuum we integrate the line emission in a $\sim$100\AA\ interval centered on the emission line peak (core). We estimate, conservatively, the continuum uncertainties from the rms of the continuum fit. The uncertainties in the line fluxes are obtained summing in quadrature the rms errors for the determination of the total flux (conservatively assumed to be the same as the continuum fit rms error) and the uncertainty in the continuum determination. 

The flux ratio between images obtained from the cores of the emission lines are used to calculate singular isothermal sphere plus shear models (SIS + $\gamma$) of HE0047-1756 and SDSS1155+6346 with  \textit{Lensmodel} \citep{keet01}. For each system, we used the separations between the two lensed images and the lens galaxy from CASTLES astrometry. From the model for each system we obtain the convergence and shear for each image ($\kappa_{A}$, $\gamma_{A}$, $\kappa_{B}$, $\gamma_{B}$) that we use to compute microlensing magnification maps.

We follow a Bayesian procedure (see e.g. Mediavilla et al. 2011) to estimate the size of the accretion disk and its temperature profile from the microlensing data. We model the accretion disk as a Gaussian with intensity profile $I(R) \propto \exp(-R^{2}/2r^2_s$), with $r_{s}$($\lambda$) $\propto$ $ \lambda^{p} $, where $r_{s}$ is the accretion disk size and $p$ is related to the temperature profile of the disk ($p$ = 4/3, \citet{ss73} thin disk model). To estimate the likelihood, $p(\Delta m_i|r_s,p)$, of reproducing the measured microlensing amplitudes, $\Delta m_i=\Delta m(\lambda_i)$, we convolve the magnification maps for the A and B images obtained using Inverse Polygon Mapping method \citep{Med06,Med2011}. We use Gaussian sources of different sizes, $r_s$, and profile slopes, $p$, at the rest frame $\lambda$ = 2045 (HE0047-1756), and $\lambda$ = 1398 (SDSS1155+6346). The size of each map is $15\times 15$ Einstein Radii ($1000\times 1000$ pixels$^{2}$). We assume a mass fraction in stars of 10\%, and we use $1M_\odot$ microlenses. From the likelihoods, we obtained Bayesian posterior probabilities, $p(r_s,p|\Delta m_i)$, using either a linear or logarithmic prior on r$_{s}$ as in \citet{Med2011} and \citet{Mot12}, to analyze the sensitivity of our study.

\section{Results}

\subsection{HE0047-1756} \label{HE0047}

HE0047-1756 is a double system discovered by \citet{wit04} with a separation between images, A and B, of 1.43$"$ \citep{Cas}. The quasar and lens galaxy redshifts are z$_{S}$ = 1.66 and z$_{L}$ = 0.41, respectively \citep{ofek06,eigen06}.

\begin{figure}
\epsscale{1.10}
\plottwo{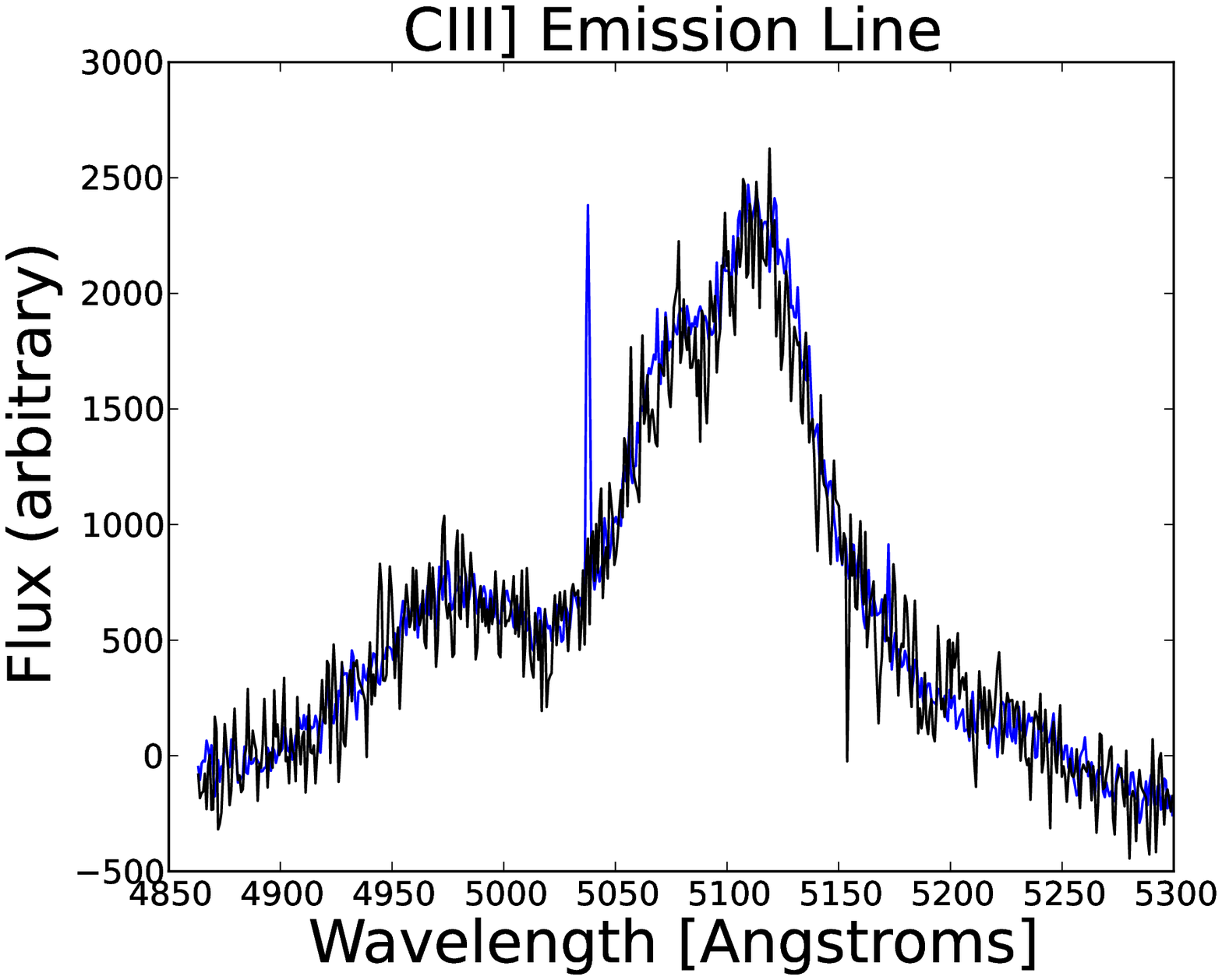}{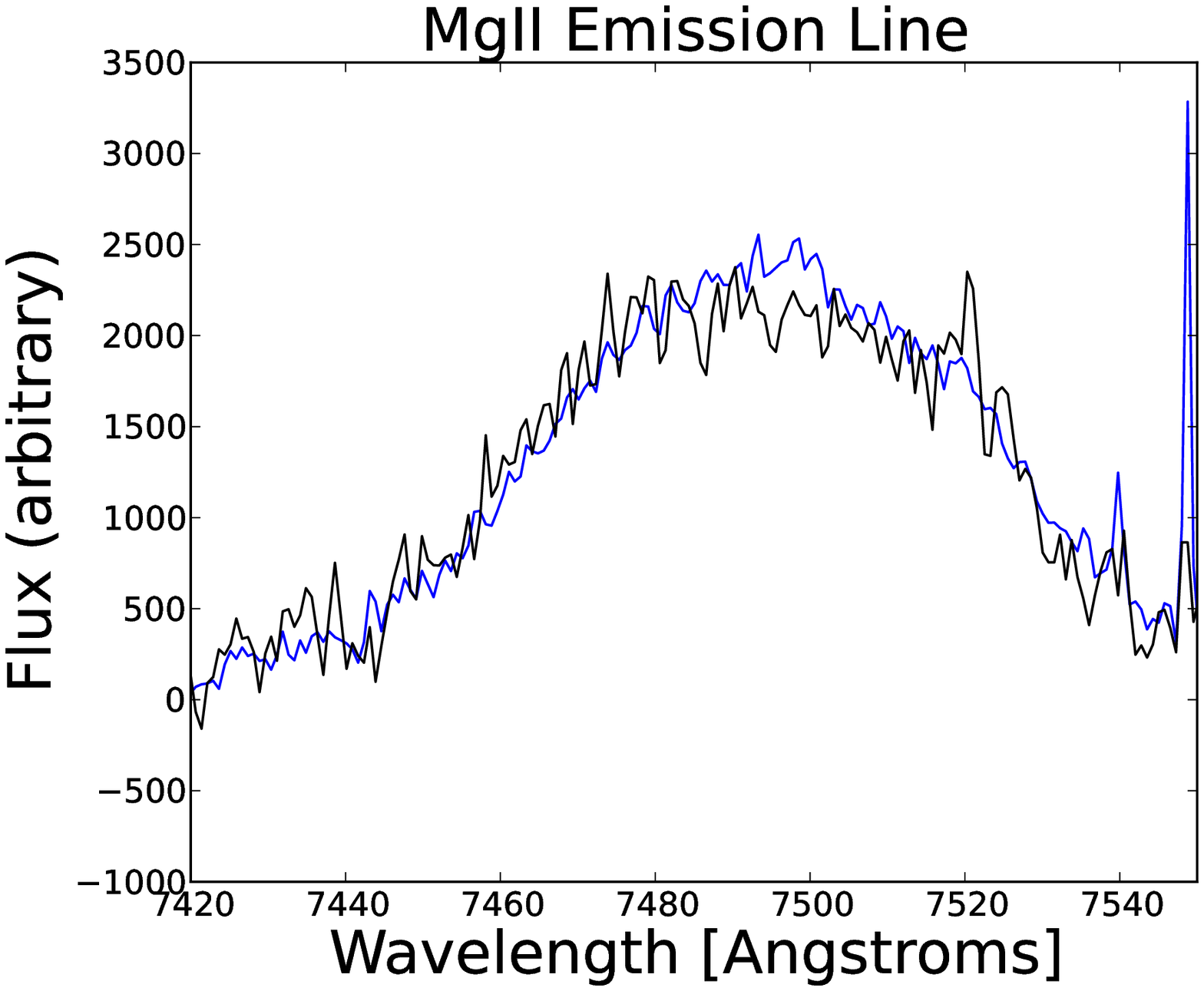}
\caption{CIII] and MgII emission lines profiles as a function of observed wavelength for HE0047-1756. The blue line is the emission line without continuum for A. The black line is the emission line without continuum for B multiplied by a factor of 4 in each case to match the peak of A.}
\label{fig1}
\end{figure}

\begin{figure}
\epsscale{.80}
\plotone{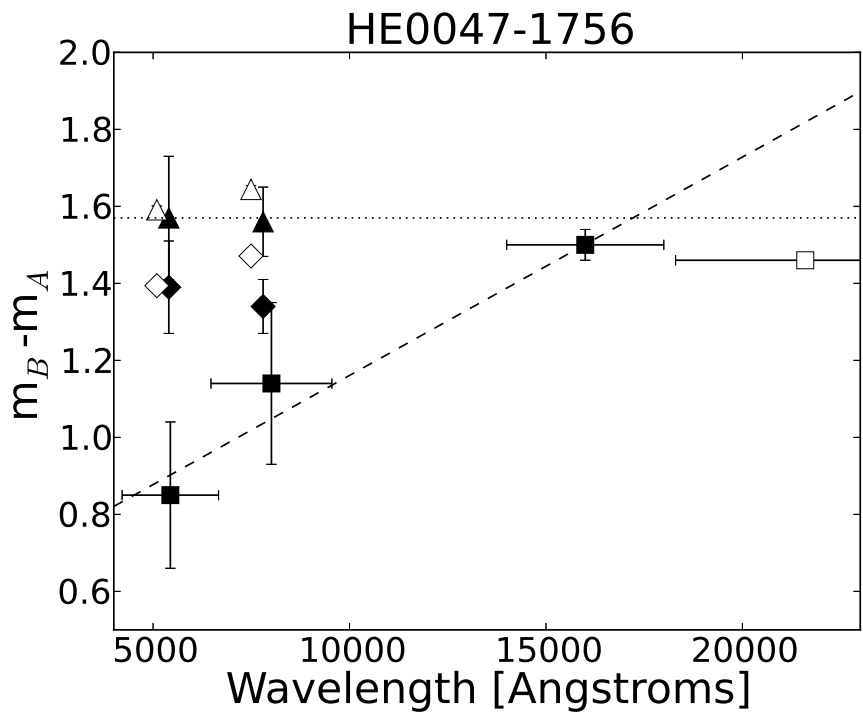}
\caption{Magnitude differences m$_{B}$ - m$_{A}$ as a function of wavelength for HE0047-1756. Solid squares are the continuum obtained from CASTLES. The horizontal error bar is the width of the band. The open square is  a K$_{s}$ band taken from \citet{wit04}. The diamonds represent magnitude differences from the continuum under the emission line core. The triangles represent the emission line cores without continuum. Open symbols are for our observed spectra and solid symbols those from \citet{slu2012}. The dashed line is the best linear fit for the CASTLES data. The dotted line is the median value for our emission lines.}
\label{fig2}
\end{figure}

In Figure \ref{fig1} the A and B spectra obtained with the Magellan telescope in 2008 are presented  in the spectral ranges corresponding to the MgII and CIII] emission lines. There are no differences between the emission line profiles that could indicate microlensing effects on the BLR. In Table \ref{tab2} and Figure \ref{fig2} we present $m_B-m_A$ magnitude differences for the emission lines and adjacent continua. There is good agreement between our results and those of \citet{slu2012}. The $m_B-m_A$ emission line average is 1.59 $\pm$ 0.02 mag. In the 2008 (this work) and 2005 \citep{slu2012} epochs there is a relatively small offset between lines and continuum that indicates microlensing of amplitude less than $0.2$ mag. 

Much more significant variations of the continuum (Fig. \ref{fig2}) are found by comparing with CASTLES data in the F555W, and F814W filters. The result for the H-band filter ($m_B-m_A$= 1.5 $\pm$ 0.04), however, agrees with the $m_B-m_A$ emission line average indicating that the region generating the emission in the H band is not affected by microlensing. The dependence with wavelength is evidence of chromatic microlensing in the 2003 epoch when the HST data were taken. The differences between the average value of the emission lines and the three CASTLES points, $\Delta m=(m_B-m_A)_{cont.}-(m_B-m_A)_{lines}$, are listed in Table \ref{tab3}.

Following the method described in \S 3 we use these wavelength dependent microlensing measurements to estimate the size and temperature profile of the accretion disk. We have used \textit{Lensmodel} \citep{keet01} to fit a SIS + $\gamma$ lens model to the images coordinates of HE0047-1756 from CASTLES and to the emission-line average flux ratio that we measured. The best fit yields a mass scale of $b$=0.75, shear of $\gamma$=0.05 and shear position $\theta_{\gamma}$=-6.44$°$ (see Table \ref{tab7}). These results are in agreement with \citet{Med09} and  \citet{slu2012}.

\begin{figure}
\epsscale{1.00}
\plottwo{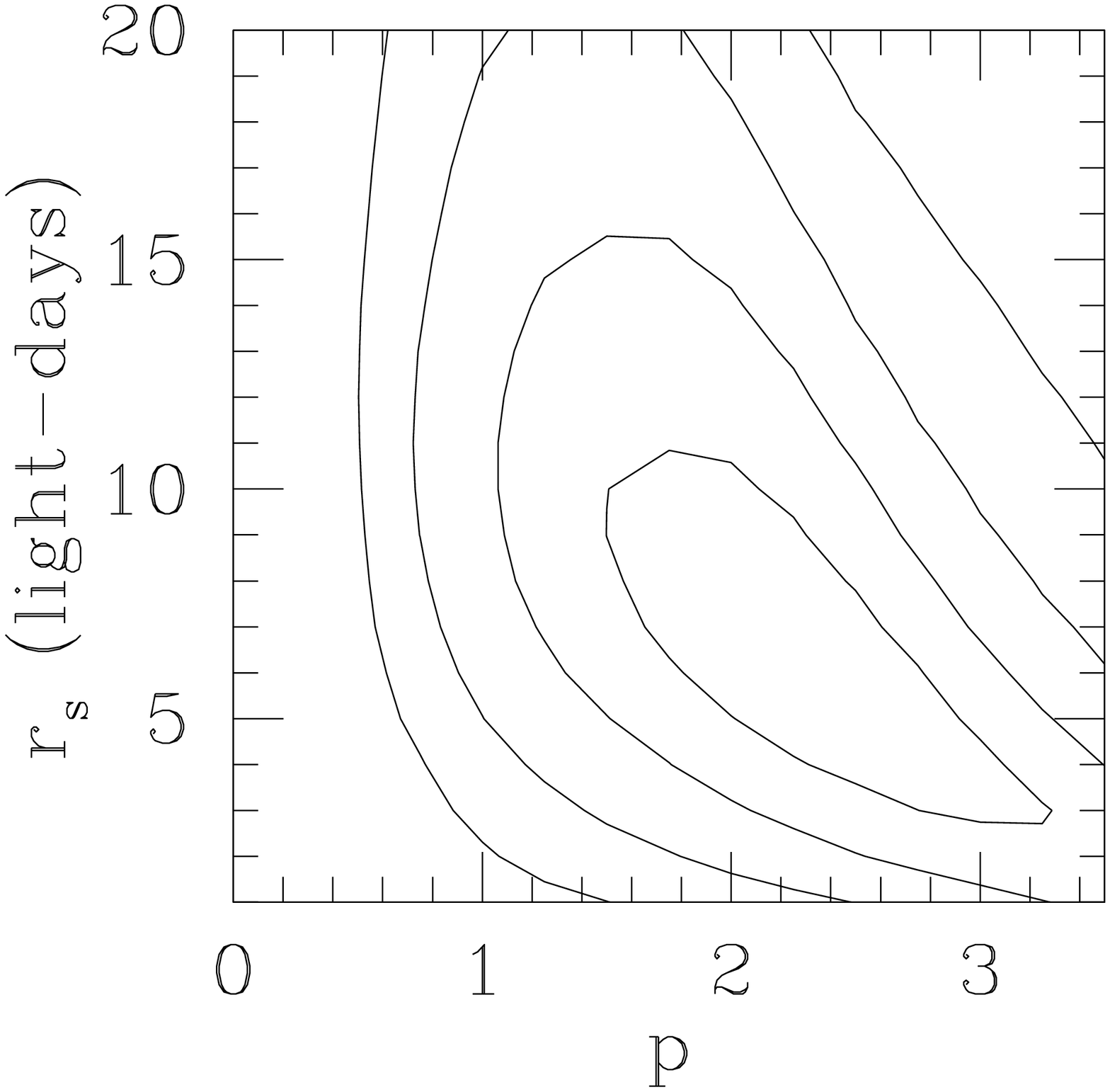}{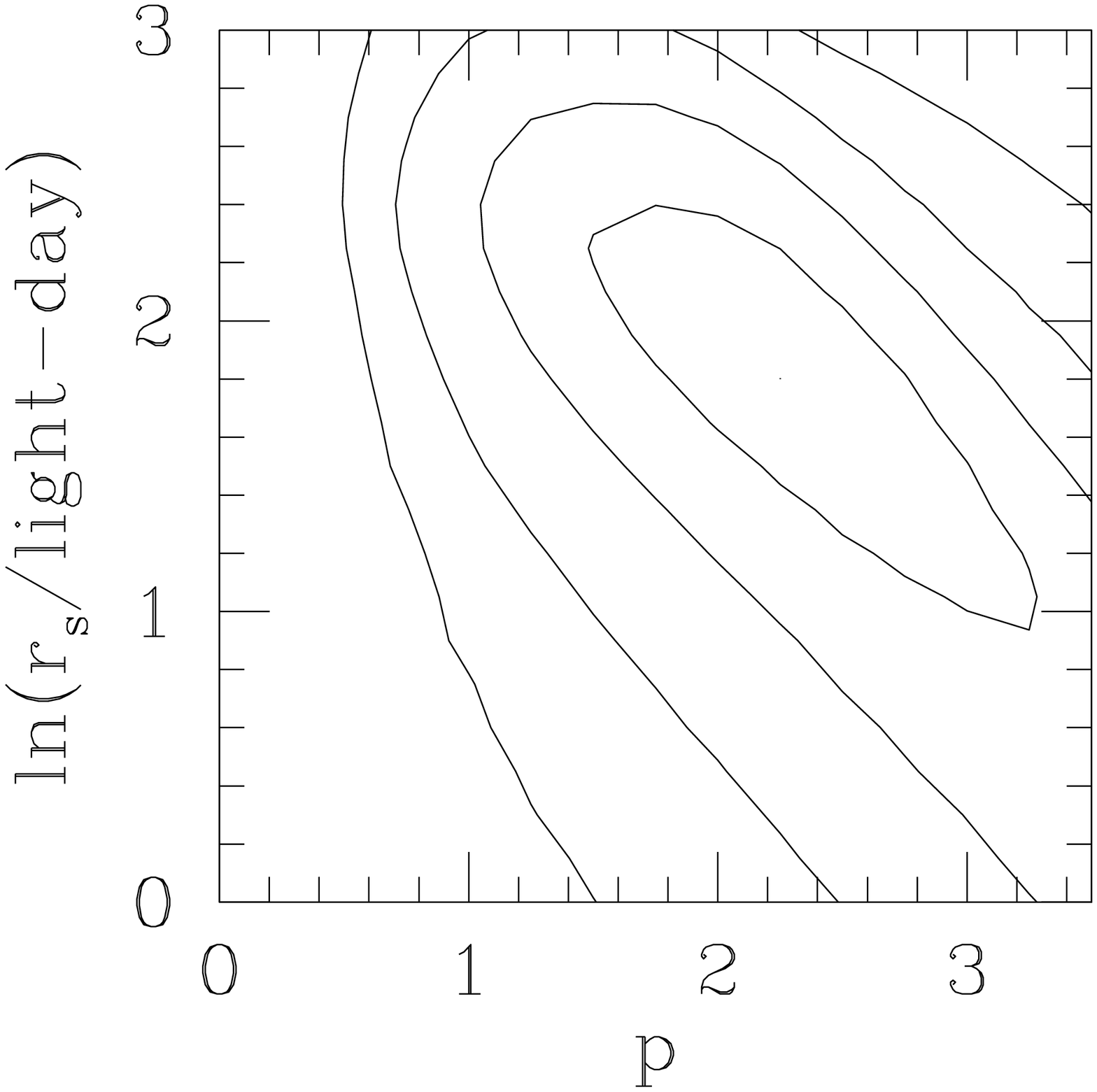}
\caption{Probability density functions for the linear size priors (left) and logarithmic size priors (right). The contours of probability are scaled in $0.5 \sigma$ steps from the maximum.}
\label{fig4}
\end{figure}

In Figure \ref{fig4} we present the probability density functions of $r_s$ and $p$ conditioned to the microlensing measurements, $\Delta m_i$ (Table \ref{tab3}), $p(r_s,p|\Delta m_i)$, using either a linear or log prior. From these probability distributions we obtain the following estimates for the accretion disk parameters: $r_{s}$ = (9.2 $\pm$ 5.0) $\sqrt{M/M_{\odot}}$ light days and p = (2.0 $\pm$ 0.8) using linear priors, and r$_{s}$ = 4.6$_{-2.5}^{+5.5}$ $\sqrt{M/M_{\odot}}$ light days and p = (2.3 $\pm$ 0.8) using logarithmic priors (Fig. \ref{fig4}). For 0.3M$_{\odot}$ microlenses the sizes would be: $r_{s}$ = (5.0 $\pm$ 2.7) $\sqrt{M/0.3M_{\odot}}$  light days (lin) and r$_{s}$ = 2.5$_{-1.4}^{+3.0}$ $\sqrt{M/0.3M_{\odot}}$  light days (log). The values for $r_{s}$ are in reasonable agreement with typical size estimates derived for other systems using microlensing (see \citet{jv2012} and references therein). Due to the large microlensing chromaticity detected in this system, the values for $p$ are larger than those predicted by \citet{ss73}, although consistent within errors. This is notable because in previous studies \citep{jv2014} microlensing chromaticity was relatively weak and the inferred values for $p$ significantly smaller than the predictions of the thin disk model.

\subsection{SDSS1155+6346} \label{SDSS1155}

SDSS1155+6346 is a double system discovered by \citet{pin04} in the Sloan Digital Sky Survey data set \citep{york2000} with a separation of 1.94$"$ between the images \citep{Cas}. \citet{pin04} measured redshifts of z$_{L} $ = 0.18 and z$_{S}$ = 2.89 for the lens and the source, respectively. The B image is within 0.2$"$ from the galaxy center and, unusually, it is the brighter component.

\begin{figure}
\epsscale{.80}
\plotone{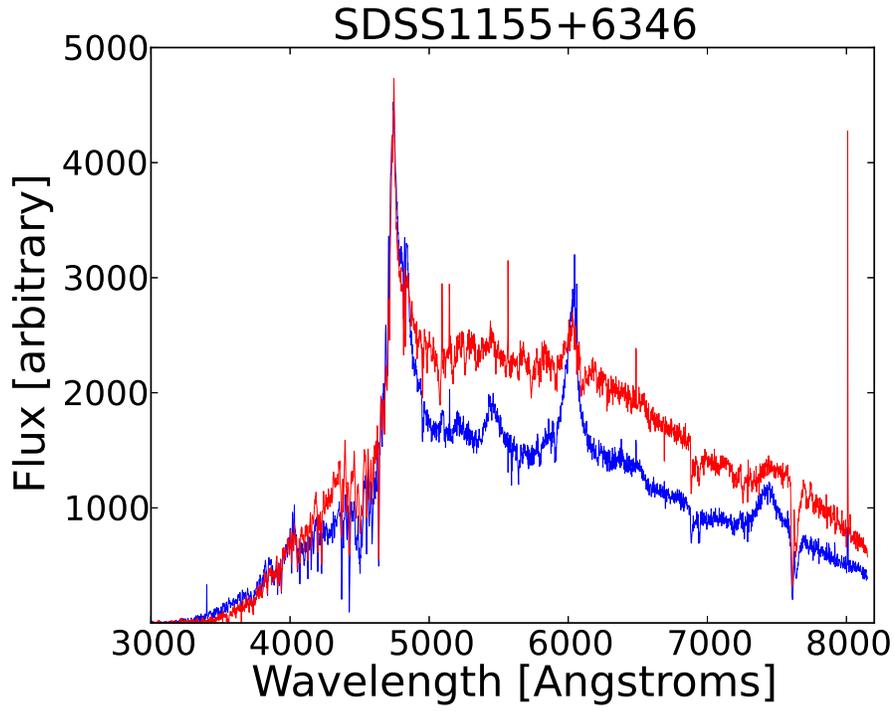}
\caption{SDSS1155+6346 spectra from the 2010 MMT observations. The A component (B) is shown in blue (red). The shapes of these spectra are very similar to those spectra obtained by Pindor et al. (2004)}
\label{sp1155}
\end{figure}

\begin{figure}
\epsscale{1.1}
\plottwo{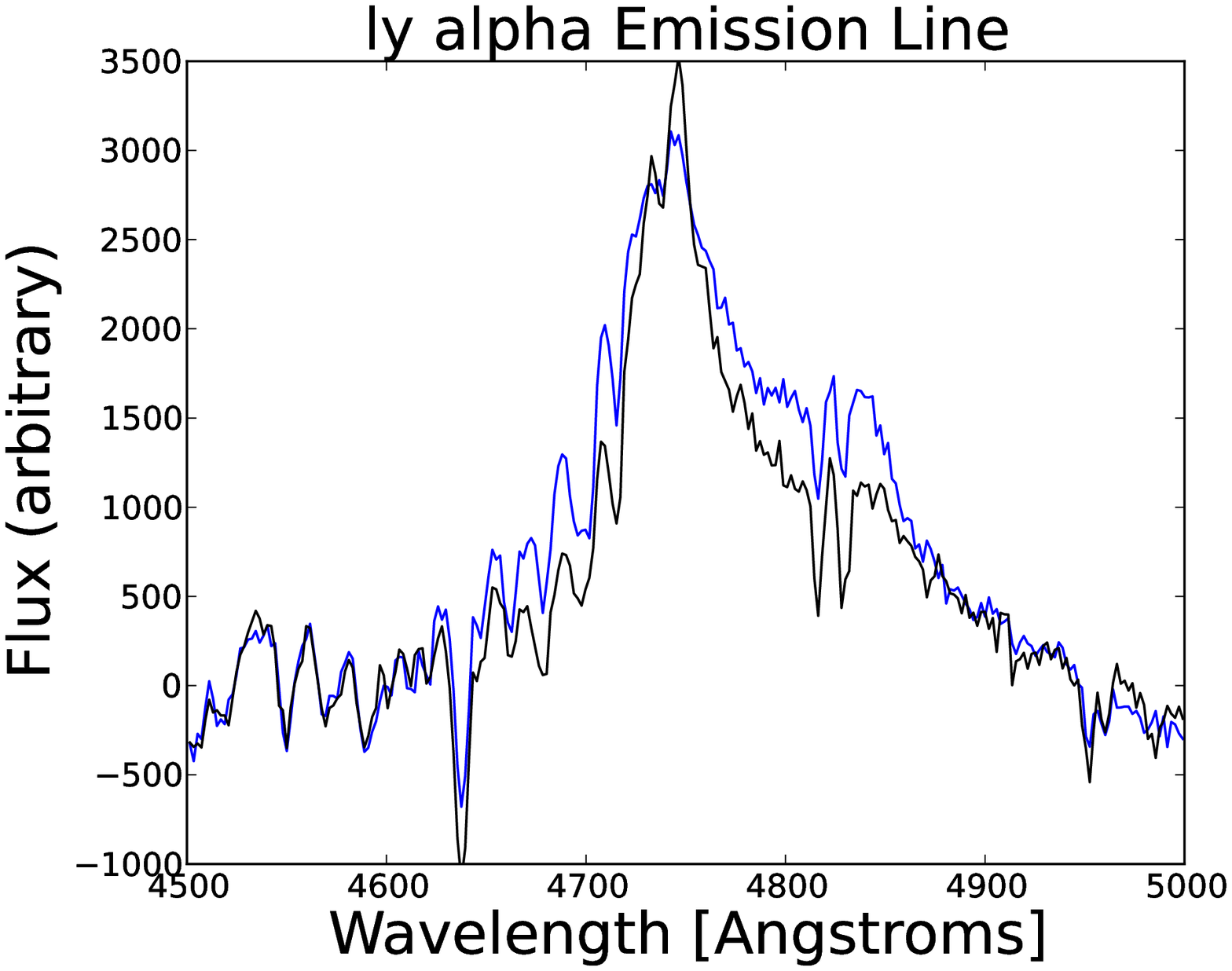}{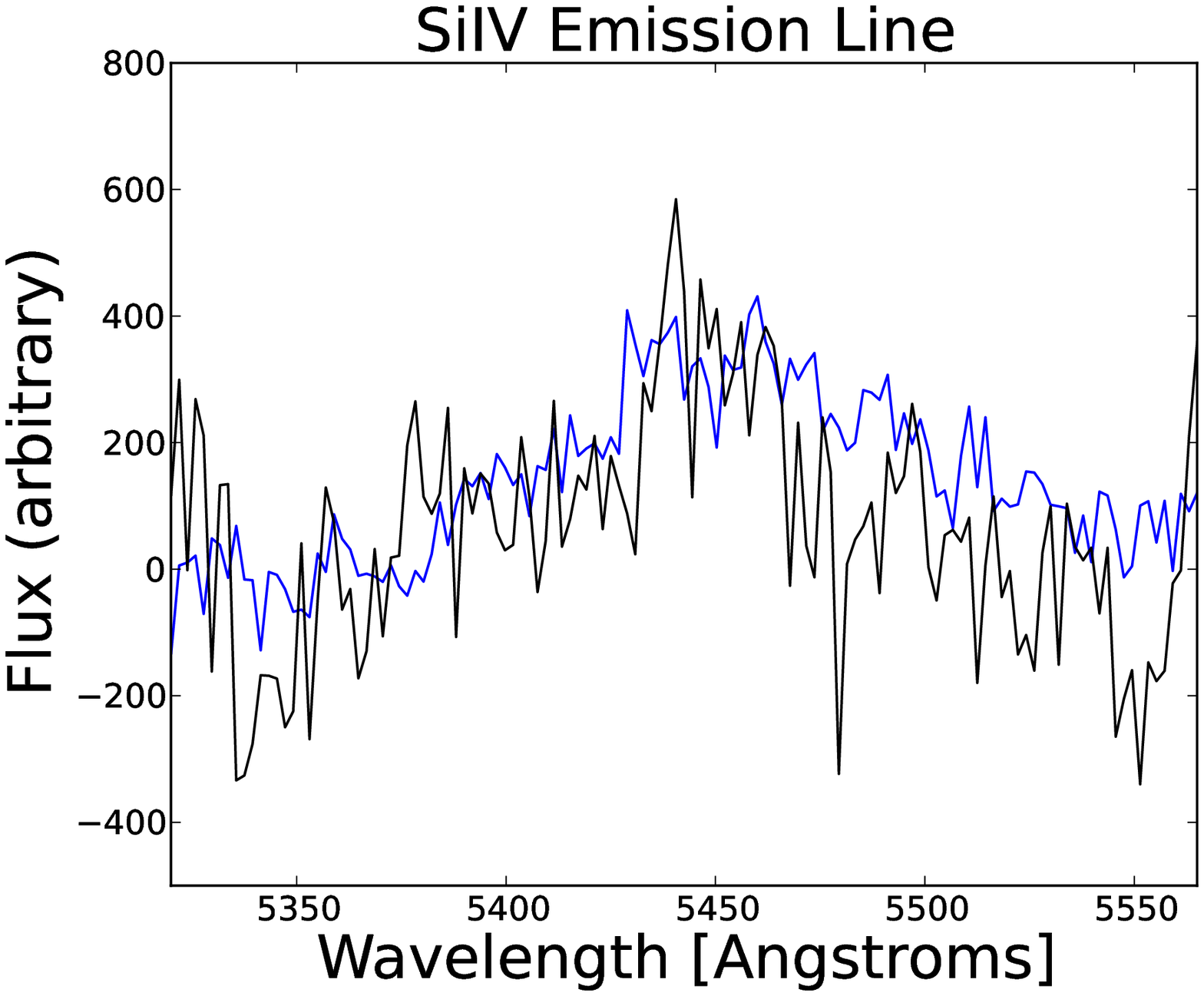}
\\
\plottwo{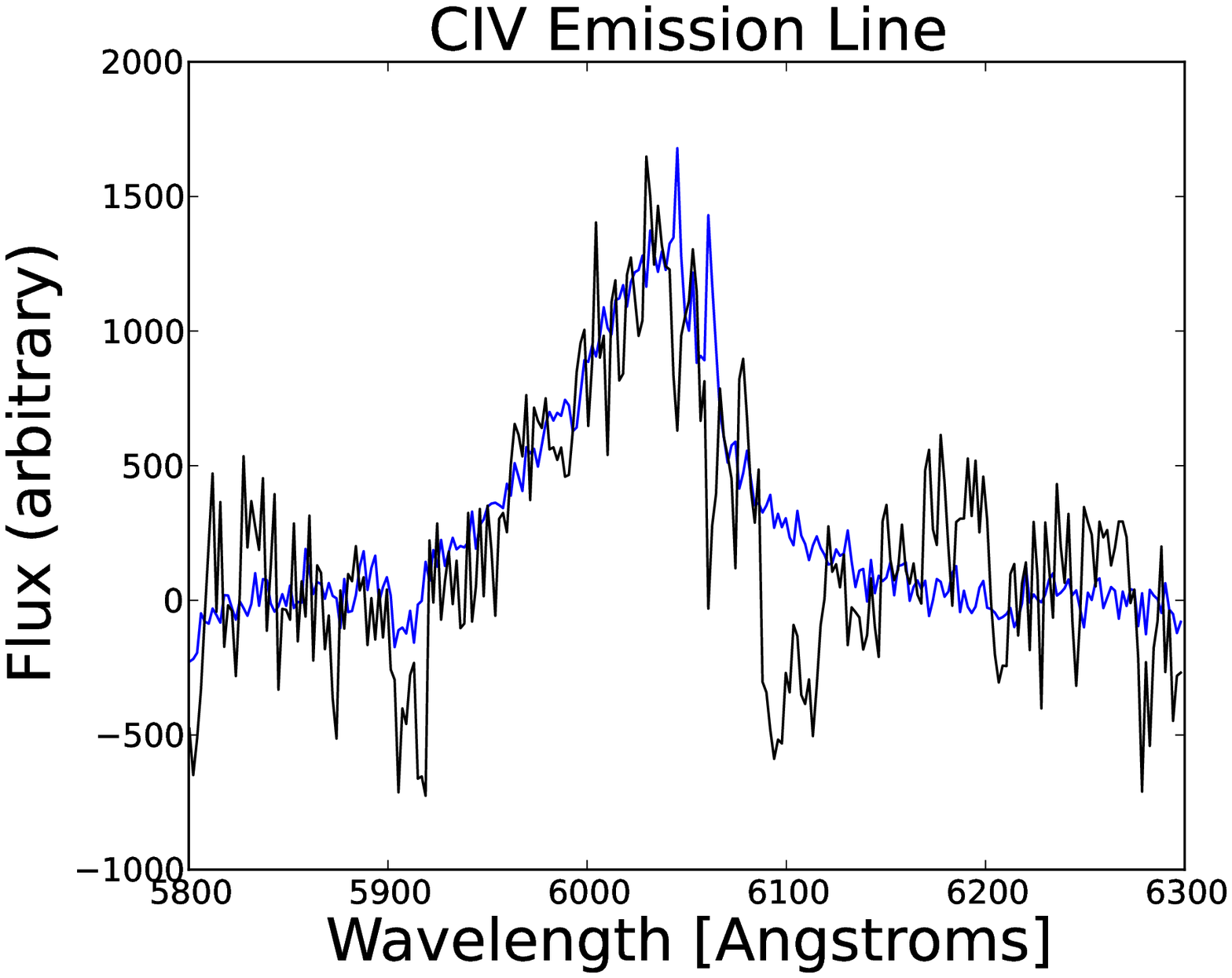}{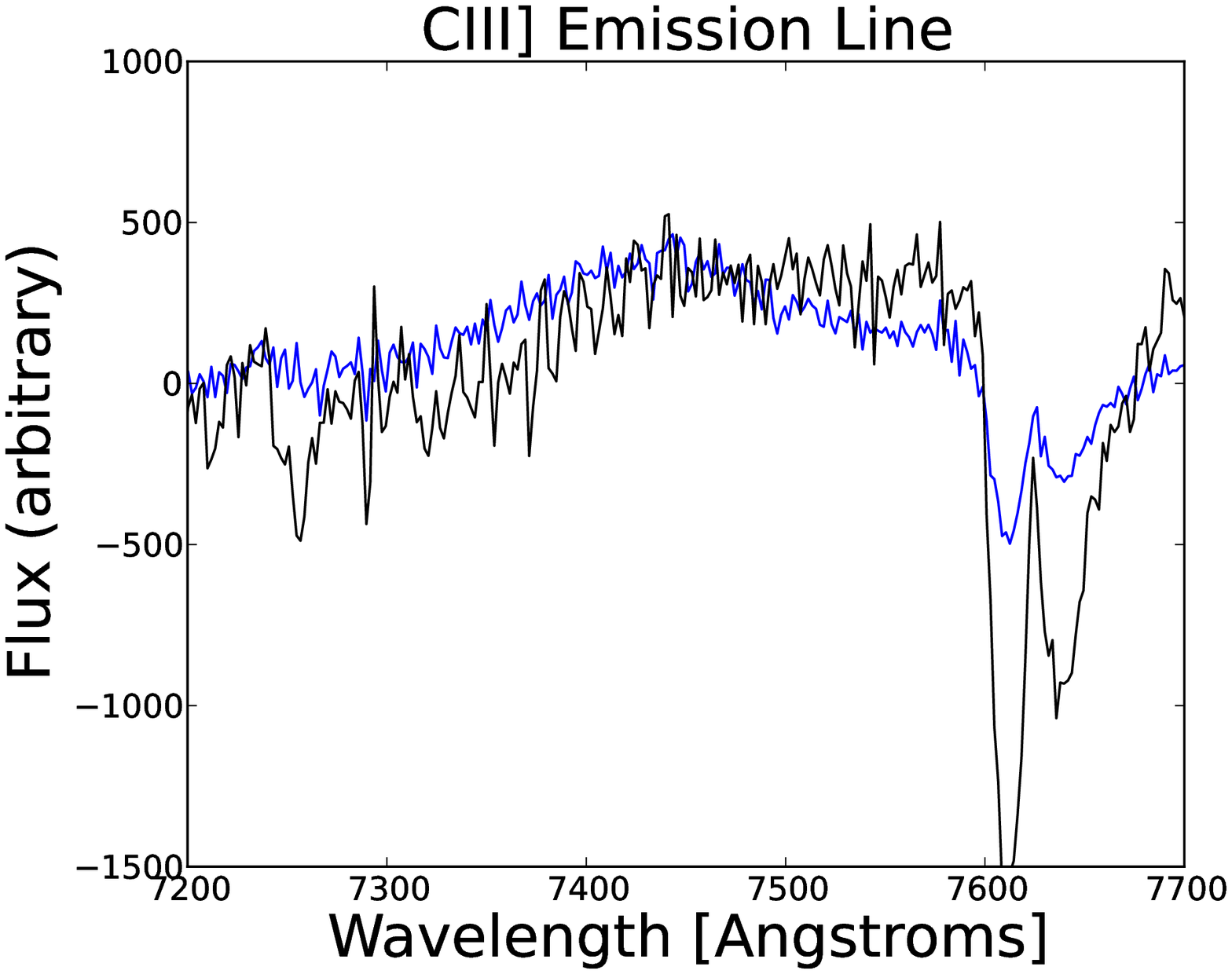}
\caption{Ly$\alpha$, SiIV, CIV, CIII] emission lines profiles as a function of observed wavelength for SDSS1155+6346. The blue line is the emission line without continuum for A. The black line is the emission line without continuum for B multiplied by a factor of 1.2 (Ly$\alpha$), 2 (SiIV), 3.2 (CIV), 2 (CIII) to match the peak of A.}
\label{fig5}
\end{figure}

\begin{figure}
\epsscale{.80}
\plotone{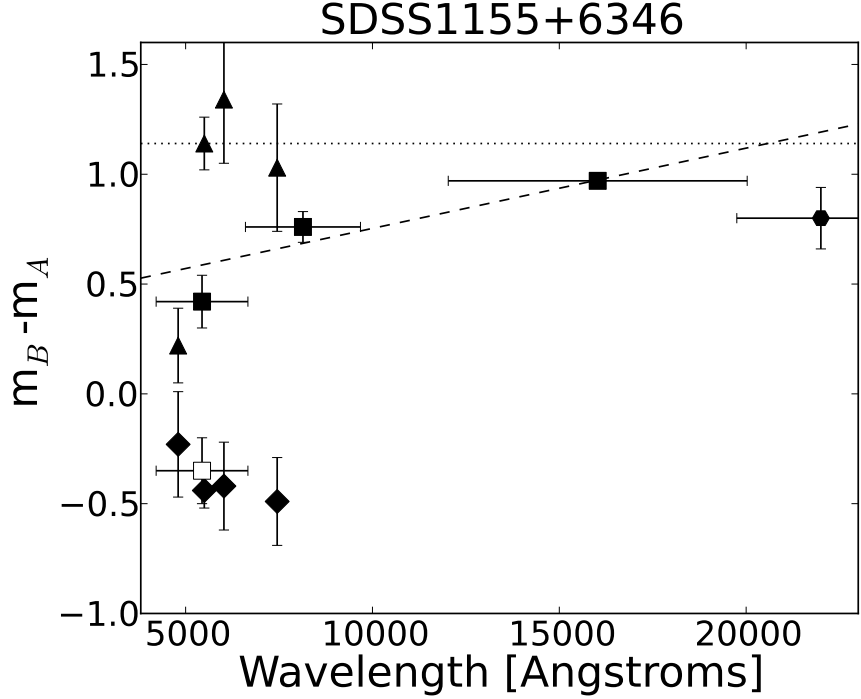}
\caption{Magnitude differences m$_{B}$ - m$_{A}$ as a function of wavelength for SDSS115+6346. The diamonds represent magnitude differences from the continuum under the emission line cores, and the triangles represent the emission line cores without continuum for our observed spectra. The dotted line is the median value for the  emission lines. The solid squares are data from CASTLES for three bands: F555W, F814W and F160W. The horizontal error bar is the width of the band. The solid hexagon is from \citet{pin04}. The dashed line is the best linear fit for the CASTLES points. The open square is the CASTLES continuum taking into account contamination from the lens galaxy.}
\label{fig6}
\end{figure}

In Figure \ref{sp1155} we present the A and B spectra from the 2010 MMT observations. These spectra are very similar in shape to those obtained by Pindor et al. (2004) if one takes into account that our data have not been flux-calibrated. The contribution from the lens galaxy to the continuum almost disappears blue-ward from Ly$\alpha$. In Figure \ref{fig5} we present the continuum-subtracted and normalized spectra in the regions corresponding to the Ly$\alpha$, SiIV, CIV, and CIII] emission lines. The A and B spectra are well matched for SiIV, CIV, and CIII] taking into account the presence of absorption features corresponding to the lens galaxy in the B spectrum. In Ly$\alpha$, however, there is a significant difference between the shapes of the line profiles corresponding to A and B images. We have tried a second order polynomial fit to the continuum and obtained the same results. These differences seem to be also present in the spectra taken by Pindor et al. (2004). However, the lens galaxy contribution to the continuum of the B image spectrum drastically changes from the red to the blue sides of Ly$\alpha$ making more uncertain the continuum subtraction and a sharp decay of the lens galaxy contribution in the red wing of Ly$\alpha$ may explain the observed differences. On the other hand, the $m_B-m_A$ magnitude differences obtained from the continua adjacent to the emission lines show a significant variation at Ly$\alpha$: -0.23 $\pm$ 0.17 mag (Ly$\alpha$), -0.44 $\pm$ 0.08 mag (SiIV region), -0.42 $\pm$ 0.20 mag (CIV region) and -0.49 $\pm$ 0.20 mag (CIII] region).  In Figure \ref{fig6} we plot the magnitude differences corresponding to the emission lines and adjacent continua with data corresponding to the F555W, F814W,  F160W, (CASTLES) and K bands (Pindor et al. 2004) obtained after subtracting the lens galaxy. The contamination from the lens galaxy is clearly present in our continuum data. In fact, if we use the F555W data without removing the contamination of the galaxy (Fig. \ref{fig6}) the resulting magnitude difference is in agreement with our data. If we leave aside the Ly$\alpha$ data that may be most contaminated by the lens galaxy continuum, the $m_B-m_A$ magnitude differences obtained from the other lines agree within the uncertainties and are also consistent with the K band data from Pindor et al. (2004), indicating that no strong differential extinction is affecting the flux ratios. If we take the average of the $m_B-m_A$ values corresponding to SiIV, CIV, and CIII] emission lines as the no microlensing baseline, $\langle m_B-m_A\rangle_{lines}=$ 1.17 $\pm$ 0.11 mag, we can determine the chromatic variation of the CASTLES continuum that will be used to estimate the size and temperature of the quasar disk (see Table \ref{tablamic}).

We use \textit{Lensmodel} \citep{keet01} to fit a SIS + $\gamma$ lens model to the image positions of SDSS1155+6346 from CASTLES and of the average flux ratio  emission lines measured (excluding Ly$\alpha$). The best fit yields a mass scale of $b$=0.78 and a very high shear of $\gamma$=0.21 with $\theta_{\gamma}$=6.63$°$ (Table \ref{tab7}). \citet{cha2010} suggest a nearby cluster may explain the high shear and ellipticity that we measured. We identify this cluster as MaxBCG J178.81693+63.83446 \citep{koe07}.

\begin{figure}
\epsscale{1.00}
\plottwo{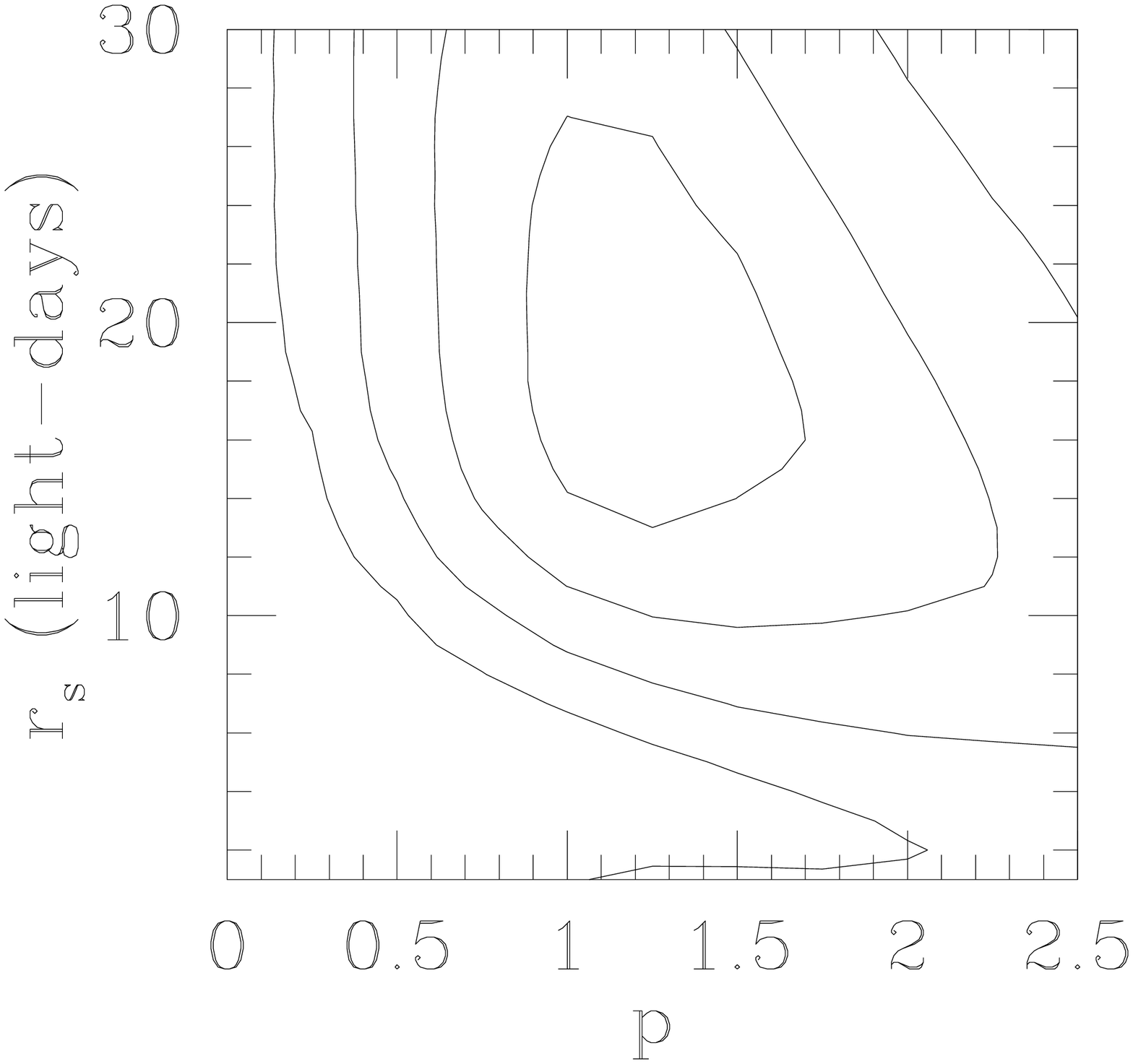}{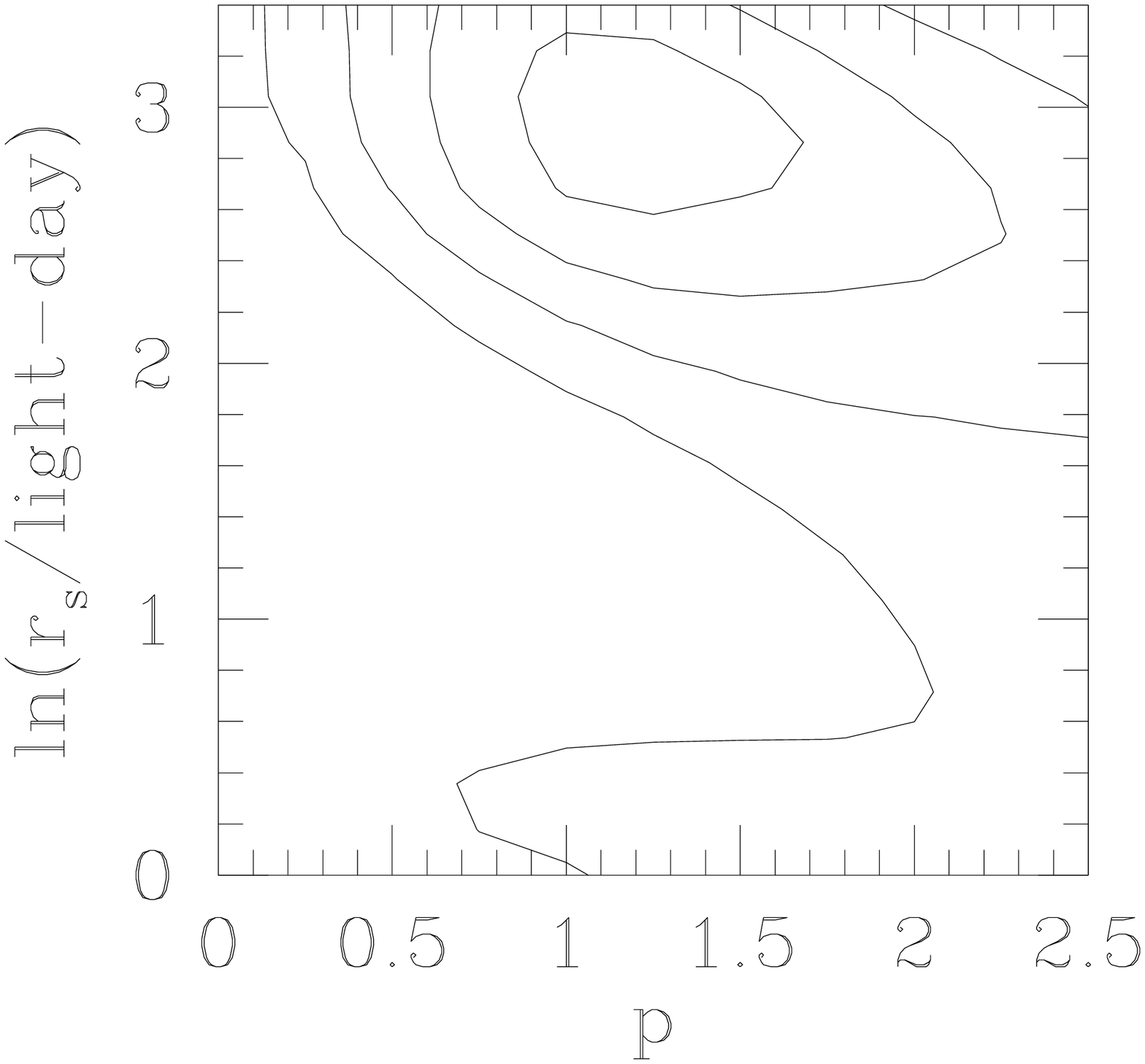}
\caption{Probability density functions for the linear size priors (left) and logarithmic size priors (right). The contours of probability are scaled in $0.5 \sigma$ steps from the maximum.}
\label{fig8}
\end{figure}

In Figure \ref{fig8} we present $p(r_s,p|\Delta m_i)$, the pdf of $r_s$ and $p$ conditioned to the microlensing measurements, $\Delta m_i$ (Table \ref{tab6} ), using either a linear or log prior. From these probability distributions we obtain the following estimates for the accretion disk parameters: r$_{s}$ = (18 $\pm$ 7) $\sqrt{M/M_{\odot}}$ light days and p = 1.4 $\pm$ 0.6 for the linear prior, and r$_{s}$ = 10$_{-6}^{+15}$ $\sqrt{M/M_{\odot}}$ light days and p = 1.5 $\pm$ 0.6 for the logarithmic prior.  For 0.3M$_{\odot}$ microlenses the sizes would be r$_{s}$ = (0.9 $\pm$ 3.8) $\sqrt{M/0.3M_{\odot}}$ light days (lin prior) and  r$_{s}$ = 5.5$_{-3.3}^{+8.2}$ $\sqrt{M/0.3M_{\odot}}$ light days (log prior).

As in the case of HE0047-1756 the large measured microlensing chromaticity implies values of $p$ consistent with the thin disk model. The inferred size is large not only compared with the thin disk model predictions but also with microlensing based estimates obtained for other lensed systems 

\section{Conclusions}

In this paper we analyzed spectroscopic data for HE0047-1756 and SDSS1155+6346 to determine the influence of microlensing and study the inner quasar structure. We point out the following results: 

1 - The shapes of the emission line profiles corresponding to the A and B images match well except in the case of Ly$\alpha$ for SDSS1155+6346, which shows strong differences in shape and an anomalous B/A flux ratio. However, the contamination from the lens galaxy in the image B spectrum strongly falls off just below this emission line, so the continuum subtraction is uncertain.

2 - When we compare the continuum from CASTLES broad band data (2003), with the no microlensing baseline consistently established for each lensed system using the emission line core flux ratios,  we find strong chromatic microlensing in both systems. In HE0047-1756 we measure microlensing amplitudes of: -0.75 $\pm$ 0.19 mag ($\lambda$5439), -0.45 $\pm$ 0.22 mag ($\lambda$8012), and -0.09 $\pm$ 0.04 mag ($\lambda$16000). In SDSS1155+6346 we measure: -0.75 $\pm$ 0.16 mag ($\lambda$5439), -0.41 $\pm$ 0.13 mag ($\lambda$8012), and -0.20 $\pm$ 0.11 mag ($\lambda$16000). 

3 - Using a Bayesian analysis, we estimate the size, $r_{s}$, and the slope of the size scaling with wavelength, $p$, of the quasar continuum emitting regions. For HE0047-1756 we found $r_{s}$ = (5.0$\pm$ 2.7) $\sqrt{M/0.3M_{\odot}}$ light days and p = 2.0 $\pm$ 0.8 (linear prior), and r$_{s}$ = 2.5$_{-1.4}^{+3.0}$ $\sqrt{M/0.3M_{\odot}}$ light days and p = 2.3 $\pm$ 0.8 (log prior). For SDSS1155+6346 we found r$_{s}$ = (9.9 $\pm$ 3.8) $\sqrt{M/0.3M_{\odot}}$ light days and p = 1.4 $\pm$ 0.6 (linear prior), and  r$_{s}$ = 5.5$_{-3.3}^{+8.2}$ $\sqrt{M/0.3M_{\odot}}$ light days and p = 1.5 $\pm$ 0.6 (log prior). The estimated values for $p$ are consistent, within errors, with the predictions of the thin disk theory but $r_s$ values are substantially larger than expected (see \cite{jv2012,jv2014}, and references therein). 

4 - Using the extinction-free and microlensing-free emission line ratios, we have computed SIS + $\gamma$ models for the two lens systems. In the SDSS1155+6346 case we found a high shear as previously found by \citet{cha2010}, which can be explained by the presence of the cluster MaxBCG J178.81693+63.83446 \citep{koe07}.

\acknowledgments

We thank the anonymous referee for thoughtful suggestions. KR and VM acknowledge support from FONDECYT through grant 1120741. KR also is supported by Doctoral scholarship FIB-UV 2014. JJV is supported by the Spanish Ministerio the Econom\'ia through grant AYA2011-24728 and by the Junta de Andaluc\'ia through project FQM-108. EM and JAM were supported by the Spanish MINECO with the grants AYA2010-21741-C03-01 and AYA2010-21741-C03-02. JAM was also supported by the Generalitat Valenciana with the project PROMETEOII/2014/060.

\clearpage

\clearpage

\clearpage

\clearpage

\begin{table}
\begin{center}
\caption{Log of observations details.}
\label{tab1}
\begin{tabular}{crrrrrr}
\tableline\tableline
System & $\Delta$\tablenotemark{a} ($"$) & Instrument & Date & Seeing ($"$)& Exposure\tablenotemark{b} & P.A.\tablenotemark{c} \\
\tableline
HE0047-1756 & 1.43 & Mag/IMACS Long Cam. & 2008/01/13  & 0.6 & 1200 & -62.9  \\
SDSS1155+6346 & 1.94 & MMT/Blue Channel & 2010/09/20 & 0.7 & 1800 & 124.9 \\

\tableline
\end{tabular}
\tablenotetext{a}{Separation between images}
\tablenotetext{b}{Seconds}
\tablenotetext{c}{Position angle in degrees E of N}
\end{center}
\end{table}

\clearpage

\begin{table}
\begin{center}
\caption{HE0047-1756 Magnitude Differences.}
\label{tab2}
\begin{tabular}{crrrrr}
\tableline\tableline
Region & $ \lambda ($ \AA $) $ & Window\tablenotemark{a} (\AA) & m$_{B}$ - m$_{A}$ (mag) & m$_{B}$ - m$_{A}$\tablenotemark{b} (mag)\\
\tableline
Continuum & 5077  & 4800-5400 & 1.39 $\pm$ 0.12 & 1.394 $\pm$ 0.003\\
		  & 7445  & 7330-7750 & 1.34 $\pm$ 0.07 & 1.471 $\pm$ 0.003\\
\tableline
Line      & CIII] & 5080-5140 & 1.57 $\pm$ 0.16 & 1.591 $\pm$ 0.007  \\
	      & MgII  & 7465-7525 & 1.56 $\pm$ 0.09 & 1.644 $\pm$ 0.005 \\
\tableline
\end{tabular}
\tablenotetext{a}{Integration window}
\tablenotetext{b}{Data from \cite{slu2012}}
\end{center}
\end{table}

\clearpage

\begin{table}
\begin{center}
\caption{HE0047-1756 Chromatic Microlensing.}
\label{tab3}
\begin{tabular}{crr}
\tableline\tableline
$ \lambda ($ \AA $) $ & $ \Delta$m$_{C}$ - $ \Delta$m$_{L}$ (mag)\\
\tableline
5439 &  -0.75 $ \pm $ 0.19  \\
8012 &  -0.45 $ \pm $ 0.22  \\
16000 & -0.09 $ \pm $ 0.04 \\
\tableline
\end{tabular}
\end{center}
\end{table}

\clearpage


\clearpage

\begin{table}
\begin{center}
\caption{SDSS1155+6346 Magnitude Differences.}
\label{tab5}
\begin{tabular}{crrr}
\tableline\tableline
Region & $ \lambda ($ \AA $) $ & Window\tablenotemark{a} (\AA) & m$_{B}$ - m$_{A}$ (mag) \\
\tableline
Continuum & 4730 & 4500-5050 & -0.23 $\pm$ 0.17 \\
		  & 5434 & 5350-5700 & -0.44 $\pm$ 0.08 \\
		  & 6025 & 5600-6400 & -0.42 $\pm$ 0.20 \\
		  & 7426 & 7000-7800 & -0.49 $\pm$ 0.20 \\
\tableline
Line & Ly$\alpha$ & 4718-4768 & 0.22 $\pm$ 0.34 \\
	 & SiIV       & 5400-5500 & 1.14 $\pm$ 0.12 \\
	 & CIV  		  & 6015-6065 & 1.34 $\pm$ 0.29 \\
	 & CIII]      & 7390-7490 & 1.03 $\pm$ 0.29 \\
\tableline
\end{tabular}
\end{center}
\end{table}

\clearpage

\begin{table}
\begin{center}
\caption{SDSS1155+6346 CASTLES continuum.}
\label{tablamic}
\begin{tabular}{crr}
\tableline\tableline
$ \lambda ($ \AA $) $ & Continuum (mag)\\
\tableline
5439  & 0.42 $ \pm $ 0.12 \\
8012  & 0.76 $ \pm $ 0.07 \\
15500 & 0.97 $ \pm $ 0.03 \\
\tableline
\end{tabular}
\end{center}
\end{table}

\clearpage

\begin{table}
\begin{center}
\caption{SDSS1155+6346 Chromatic Microlensing.}
\label{tab6}
\begin{tabular}{crr}
\tableline\tableline
$ \lambda ($ \AA $) $ & $ \Delta$m$_{C}$ - $ \Delta$m$_{L}$ (mag)\\
\tableline
5439  & -0.75 $ \pm $ 0.16 \\
8012  & -0.41 $ \pm $ 0.13 \\
15500 & -0.20 $ \pm $ 0.11 \\
\tableline
\end{tabular}
\end{center}
\end{table}

\clearpage

\begin{table}
\begin{center}
\caption{Results from \textit{Lensmodel}.}
\label{tab7}
\begin{tabular}{ccrrrrrrrrrrr}
\tableline\tableline
System & Model  & b($"$) & $\gamma$ & $\theta_{\gamma}$ & f$_{b}$/f$_{a}$ & $ \kappa_{A} $& $ \gamma_{A} $ & $ \kappa_{B}$ & $ \gamma_{B} $ \\
\tableline
HE0047-1756  & SIS+$\gamma$ & 0.75   & 0.05 & -6.44 & 0.24 & 0.45 & 0.48 & 0.62 & 0.66\\
SDSS1155+6346 & SIS+$\gamma$ & 0.78 & 0.21 & 6.63 & 0.34 & 0.22 & 0.03 & 1.67 & 1.47\\
\tableline
\end{tabular}
\end{center}
\end{table}


\end{document}